# Yang-Baxter algebras based on the two-colour BWM algebra

Uwe Grimm[*] and S. Ole Warnaar[†]

June 1995

**Abstract**

We present a Baxterization of a two-colour generalization of the Birman–Wenzl–Murakami (BWM) algebra. Appropriately combining two RSOS-type representations of the ordinary BWM algebra, we construct representations of the two-colour algebra. Using the Baxterization, this provides new RSOS-type solutions to the Yang–Baxter equation.

## 1 Introduction

Since the work of Baxter [1], the full relevance of the Yang–Baxter equation (YBE) in the theory of two-dimensional solvable lattice models has been realized. Among the several algebraic techniques used to construct solutions to the YBE, a particularly interesting approach is based on braid-monoid algebras [2]. This method amounts to reducing the problem of finding representations of the Yang–Baxter algebra (YBA) to that of finding representations of certain types of braid-monoid algebras, through a procedure called *Baxterization* [3]. Some examples of braid-monoid algebras for which a Baxterization is known are the Temperley–Lieb algebra [4], the Birman–Wenzl–Murakami (BWM) algebra [5, 6] and their dilute generalizations [7, 8, 9].

In Ref. [10], solvable lattice models related to a two-colour generalization of the Temperley–Lieb algebra were found. This has motivated our attempt to also find a Baxterization of a two-colour generalization of the BWM algebra.

This paper is organized as follows. We first give a short definition of the two-colour braid-monoid algebra, which, apart from a small simplification, coincides with the general definition of Ref. [11]. From this the two-colour BWM algebra is obtained by imposing polynomial reduction relations on the generators. In Sec. 3, we present a Baxterization of the two-colour BWM algebra and relate it to the dilute BWM [9, 12, 13, 14] and the two-colour Temperley–Lieb case [10]. Subsequently, in Sec. 4, we construct RSOS-type representations of the two-colour algebra. This leads to new RSOS-type representations of the Yang–Baxter algebra and hence to new solvable lattice models. Finally, we summarize and discuss our results in Sec. 5.

[*]Instituut voor Theoretische Fysica, Universiteit van Amsterdam, Valckenierstraat 65, 1018 XE Amsterdam, The Netherlands, e-mail: `grimm@phys.uva.nl`

[†]Mathematics Department, University of Melbourne, Parkville, Victoria 3052, Australia, e-mail: `warnaar@maths.mu.oz.au`

## 2  The two-colour BWM algebra

We commence by defining a *two-colour braid-monoid algebra* following Ref. [11]. It is generated by the following set of operators:

$$\begin{aligned}
&p_j^{(c,c')} &&\text{(projectors)} \\
&b_j^{\pm(c,c)},\ b_j^{(c,\bar{c})} &&\text{(braids)} \\
&e_j^{(c,c')} &&\text{(Temperley–Lieb operators)},
\end{aligned} \tag{2.1}$$

where $j = 1, 2, \ldots, N-1$ and $N$ is some number. Here $c, c' \in \{1, 2\}$ label the two colours, and $\bar{c} = 3 - c$. Furthermore, we have central elements $\sqrt{Q_c}$ and $\omega_c$ associated to each colour. The algebra is defined by imposing a number of relations discussed below. However, let us first give a graphical interpretation of the generators to motivate the defining relations.

We view the generators (2.1) as acting on a set of $N$ strings labelled by $j = 1, 2, \ldots, N$, with non-trivial action at positions $j$ and $j + 1$ only. Multiplication in the algebra corresponds to concatenation of the respective diagrams. At each position, we initially have a "white" (or "colourless") string corresponding to the identity $\mathcal{I}$ in the algebra. Hence any of the generators (2.1) selects particular colours for the strings at positions $j$ and $j + 1$. In particular, $p_j^{(c,c')}$ is interpreted as doing just this — it creates a string of colour $c$ at position $j$ and a string of colour $c'$ at position $j + 1$. More precisely, the $p_j^{(c,c')}$ are required to satisfy

$$\begin{aligned}
p_j^{(c'',c''')}\, p_j^{(c,c')} &= \delta_{c,c''}\delta_{c',c'''}\, p_j^{(c,c')} \\
\sum_{c,c'} p_j^{(c,c')} &= \mathcal{I}
\end{aligned} \tag{2.2}$$

and hence are orthogonal projectors. The non-trivial part of these and the remaining generators can be represented by the following diagrams

$$p^{(1,1)} = \rangle\langle \qquad p^{(1,2)} = \rangle\!\langle\!\langle \qquad p^{(2,1)} = \rangle\!\rangle\langle \qquad p^{(2,2)} = \rangle\!\rangle\!\langle\!\langle$$

$$e^{(1,1)} = \underset{\cap}{\cup} \qquad e^{(1,2)} = \underset{\cap}{\cup\!\!\cup} \qquad e^{(2,1)} = \underset{\cap\!\!\cap}{\cup} \qquad e^{(2,2)} = \underset{\cap\!\!\cap}{\cup\!\!\cup}$$

$$b^{+\,(1,1)} = \times \qquad b^{-\,(1,1)} = \times \qquad b^{+\,(2,2)} = \times\!\!\!\times \qquad b^{-\,(2,2)} = \times\!\!\!\times$$

$$b^{(1,2)} = \times \qquad b^{(2,1)} = \times \qquad . \tag{2.3}$$

Note that our set of generators is in fact smaller than that of Ref. [11] since we do not distinguish between over- and undercrossing for strings of different colour. Nevertheless, for the sake of brevity we also use a unified notation for the braids by setting $b_j^{(c,\bar{c})} = b_j^{+\,(c,\bar{c})} = b_j^{-\,(c,\bar{c})}$ for the corresponding generators.

Now, we demand that any two diagrams which can be deformed into each other by continuous deformations of strings correspond to the same element in the algebra, apart from certain factors carried by so-called "twists" and by closed loops. In particular, any diagram where colours do not match corresponds to the zero element of the algebra. This yields a number of obvious compatibility relations between the generators (2.1) acting at the same or at neighbouring positions (see Ref. [11] for a more formal treatment). Also, any two generators acting at positions $j$ and $k$ with $|j - k| \geq 2$ trivially commute. Besides these, one has numerous relations, for which a sufficient (though not minimal) subset is given by the following three lists. The first consists of the braid relations

$$b_j^{-\,(c',c)} b_j^{+\,(c,c')} = b_j^{+\,(c',c)} b_j^{-\,(c,c')} = p_j^{(c,c')}$$
$$b_{j+1}^{+\,(c'',c)} b_j^{+\,(c'',c')} b_{j+1}^{+\,(c,c')} = b_j^{+\,(c,c')} b_{j+1}^{+\,(c'',c')} b_j^{+\,(c'',c)} \tag{2.4}$$

for the coloured braids. The second contains the Temperley–Lieb relations

$$e_j^{(c',c'')} e_j^{(c,c')} = \sqrt{Q_{c'}}\, e_j^{(c,c'')}$$
$$e_j^{(c',c'')} e_{j+1}^{(c',c')} e_j^{(c,c')} = e_j^{(c,c'')} p_{j+1}^{(c,c')}$$
$$e_j^{(c',c'')} e_{j-1}^{(c',c')} e_j^{(c,c')} = e_j^{(c,c'')} p_{j-1}^{(c',c)} \tag{2.5}$$

and finally we have the braid-monoid relations

$$b_j^{+(c',c)} e_j^{(c,c')} = \omega_{c'} e_j^{(c,c')}$$

$$e_j^{(c,c')} b_j^{+(c,c)} = \omega_c e_j^{(c,c')}$$

$$b_{j+1}^{+(c'',c')} b_j^{+(c'',c)} e_{j+1}^{(c,c')} = e_j^{(c,c')} b_{j+1}^{+(c'',c)} b_j^{+(c'',c)} = e_j^{(c'',c')} e_{j+1}^{(c,c'')}$$

$$b_{j-1}^{+(c',c'')} b_j^{+(c',c'')} e_{j-1}^{(c,c')} = e_j^{(c,c')} b_{j-1}^{+(c,c'')} b_j^{+(c,c'')} = e_j^{(c'',c')} e_{j-1}^{(c,c'')} \,. \tag{2.6}$$

Eqs. (2.4)–(2.6) are just the relations one obtains by considering all possible "2-colourings" of the diagrams corresponding to the relations defining the usual (one-colour) braid-monoid algebra [2].

Finally, the two-colour braid-monoid algebra defined above becomes a *two-colour BWM algebra* if both one-colour subalgebras are of BWM type [5, 6]. This means that the braids satisfy the cubic reduction relations

$$\left(b_j^{+(c,c)} - q_c^{-1} p_j^{(c,c)}\right)\left(b_j^{+(c,c)} + q_c p_j^{(c,c)}\right)\left(b_j^{+(c,c)} - \omega_c p_j^{(c,c)}\right) = 0 \tag{2.7}$$

and the Temperley–Lieb generators $e_j^{(c,c)}$ are given by quadratic expressions in the braids as follows

$$e_j^{(c,c)} = \frac{\omega_c^{-1}}{q_c - q_c^{-1}} \left(b_j^{+(c,c)} - q_c^{-1} p_j^{(c,c)}\right)\left(b_j^{+(c,c)} + q_c p_j^{(c,c)}\right)$$

$$= p_j^{(c,c)} + \frac{b_j^{+(c,c)} - b_j^{-(c,c)}}{q_c - q_c^{-1}} \,. \tag{2.8}$$

Here $q_c$ is related to $\sqrt{Q_c}$ and $\omega_c$ by

$$\sqrt{Q_c} = 1 + \frac{\omega_c - \omega_c^{-1}}{q_c - q_c^{-1}} \tag{2.9}$$

as follows from Eqs. (2.5), (2.7) and (2.8).

# 3 Baxterization

We are interested in constructing solutions of the YBE based on representations of the two-colour BWM algebra. To this end, we introduce local face operators $X_j(u)$, $j = 1, \ldots, N-1$, which depend on the spectral parameter $u$. We want the face operators to generate a *Yang-Baxter algebra* (YBA). That is, the face operators $X_j(u)$ satisfy the YBE

$$X_j(u) X_{j+1}(u+v) X_j(v) = X_{j+1}(v) X_j(u+v) X_{j+1}(u) \tag{3.1}$$

and the commutation relations

$$X_j(u) X_k(v) = X_k(v) X_j(u) \qquad \text{for } |j - k| \geq 2 \,. \tag{3.2}$$

## 3.1 Two-colour BWM algebra

In order to Baxterize the two-colour BWM algebra, let us consider a quotient of the two-colour algebra obtained by imposing the condition

$$q_1 = q_2 = q. \tag{3.3}$$

However, we still allow $\omega_1$ and $\omega_2$, and hence $\sqrt{Q_1}$ and $\sqrt{Q_2}$, to take different values. We introduce $\lambda$ and $\eta$ by

$$\mathrm{e}^{-\mathrm{i}\lambda} = q$$
$$\mathrm{e}^{-2\mathrm{i}\eta\lambda} = q^{2\eta} = \omega_1\omega_2. \tag{3.4}$$

Then the defining relations of the two-colour BWM algebra are sufficient to show that

$$\begin{aligned}
X_j(u) &= p_j^{(1,1)} + p_j^{(2,2)} + \frac{\sin(\eta\lambda - u)}{\sin\eta\lambda}\left(p_j^{(1,2)} + p_j^{(2,1)}\right) \\
&\quad - \frac{\sin u}{2\mathrm{i}\sin\lambda\sin\eta\lambda}\left[\mathrm{e}^{\mathrm{i}(\eta\lambda - u)}\left(b_j^{+(1,1)} + b_j^{+(2,2)}\right) - \mathrm{e}^{\mathrm{i}(u-\eta\lambda)}\left(b_j^{-(1,1)} + b_j^{-(2,2)}\right)\right] \\
&\quad + \frac{\sin u\,\sin(\eta\lambda - u)}{\sin\lambda\sin\eta\lambda}\left(b_j^{(1,2)} + b_j^{(2,1)}\right) + \frac{\sin u}{\sin\eta\lambda}\left(e_j^{(1,2)} + e_j^{(2,1)}\right)
\end{aligned} \tag{3.5}$$

satisfies the defining relations (3.1)–(3.2) of the YBA. Therefore, any representation of the two-colour BWM algebra with $q_1 = q_2$ gives rise to a solution of the YBE via Eq. (3.5). These solutions are crossing-symmetric with crossing parameter $\eta\lambda$, and satisfy the inversion relation

$$X_j(u)\,X_j(-u) = \varrho(u)\,\varrho(-u)\,\mathcal{I} \tag{3.6}$$

where the function $\varrho(u)$ takes the form

$$\varrho(u) = \frac{\sin(\lambda - u)\sin(\eta\lambda - u)}{\sin\lambda\sin\eta\lambda}. \tag{3.7}$$

## 3.2 Dilute BWM algebra

A particularly interesting simplification occurs if we set $\omega_2 = \sigma$ with $\sigma^2 = 1$. By Eq. (2.9), this implies $\sqrt{Q_2} = 1$ and thus we can sum out the degrees of freedom associated with the second colour. To be precise, we get

$$b_j^{+(2,2)} = b_j^{-(2,2)} = \sigma\,e_j^{(2,2)} = \sigma\,p_j^{(2,2)}, \tag{3.8}$$

and graphically we can represent the generators (2.3) as in Figure 1. These we recognize as the generators of the dilute BWM algebra [9, 12, 13, 14]. Using Eq. (3.8), the relations

(2.2)–(2.9) indeed reduce to those of the dilute BWM algebra, and the Baxterization (3.5) becomes

$$X_j(u) = p_j^{(1,1)} - \frac{\sin u}{2\,\mathrm{i}\,\sin\lambda\,\sin\eta\lambda}\left(\mathrm{e}^{\mathrm{i}(\eta\lambda-u)}\,b_j^{+(1,1)} - \mathrm{e}^{\mathrm{i}(u-\eta\lambda)}\,b_j^{-(1,1)}\right)$$
$$+ \frac{\sin u\,\sin(\eta\lambda-u)}{\sin\lambda\,\sin\eta\lambda}\left(b_j^{(1,2)}+b_j^{(2,1)}\right) + \frac{\sin(\eta\lambda-u)}{\sin\eta\lambda}\left(p_j^{(1,2)}+p_j^{(2,1)}\right)$$
$$+ \frac{\sin u}{\sin\eta\lambda}\left(e_j^{(1,2)}+e_j^{(2,1)}\right) + \left(1+\sigma\frac{\sin u\,\sin(\eta\lambda-u)}{\sin\lambda\,\sin\eta\lambda}\right)p_j^{(2,2)},\qquad(3.9)$$

which coincides with the Baxterization of the dilute BWM algebra of Refs. [9, 12, 13, 14].

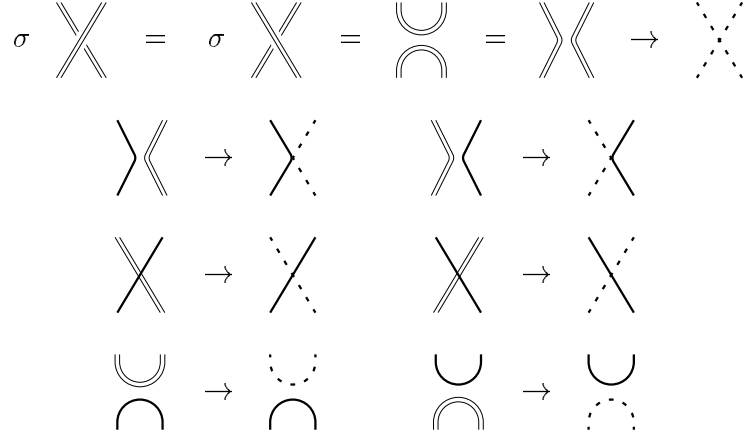

Figure 1: Graphical representation of the two-colour BWM generators after summing out the second colour. Generators which do not involve the second colour remain as in Eq. (2.3).

### 3.3 Two-colour Temperley–Lieb algebra

Another simplification worth mentioning occurs if we demand the additional reduction relation

$$e_j^{(c,c)} = \frac{1}{q^2-q^{-2}}\left(q\,b_j^{+(c,c)} - q^{-1}\,b_j^{-(c,c)}\right) \qquad(3.10)$$

for both $c=1$ and $c=2$. Combining this with Eq. (2.8) yields the following quadratic equation for the braids

$$\left(b_j^{+(c,c)} - q^{-1}\,p_j^{(c,c)}\right)\left(b_j^{+(c,c)} + q^3\,p_j^{(c,c)}\right) = 0. \qquad(3.11)$$

Comparing with the cubic (2.7), we conclude that Eq. (3.10) is consistent with the two-colour BWM algebra provided

$$\omega_1 = \omega_2 = -q^3. \qquad(3.12)$$

By Eq. (3.4), this fixes $\eta$ to be $\eta = 3$. Using both reduction relations (2.8) and (3.10), we can eliminate the braids $b_j^{\pm(c,c)}$ from Eq. (3.5) to obtain

$$X_j(u) = \frac{\sin(3\lambda - u) \sin(\lambda - u)}{\sin \lambda \sin 3\lambda} \left( p_j^{(1,1)} + p_j^{(2,2)} \right) + \frac{\sin(3\lambda - u)}{\sin 3\lambda} \left( p_j^{(1,2)} + p_j^{(2,1)} \right)$$
$$- \frac{\sin u \sin(2\lambda - u)}{\sin \lambda \sin 3\lambda} \left( e_j^{(1,1)} + e_j^{(2,2)} \right) + \frac{\sin u}{\sin 3\lambda} \left( e_j^{(1,2)} + e_j^{(2,1)} \right)$$
$$+ \frac{\sin u \sin(3\lambda - u)}{\sin \lambda \sin 3\lambda} \left( b_j^{(1,2)} + b_j^{(2,1)} \right) . \tag{3.13}$$

This is precisely the Baxterization of the two-colour Temperley–Lieb algebra of Ref. [10].

## 4 Representations of the two-colour BWM algebra

In this section we construct RSOS-type representations of the two-colour BWM algebra. Basically this amount to appropriately combining two arbitrary RSOS-type representations of the ordinary BWM algebra. For these we take the representations labelled by the $B_n^{(1)}$, $C_n^{(1)}$, and $D_n^{(1)}$ affine Lie algebras, found by Deguchi et al. [15, 2]. Our procedure is similar to the construction of representations of the dilute BWM algebra out of ordinary BWM algebra representations [12, 13, 14]. All representations presented below have $\omega_2 \neq \pm 1$. For RSOS representations of the dilute case of the two-colour BWM algebra we refer to Refs. [13, 14]. The representations given below labelled by (C,C) with $n_1 = n_2 = 1$ satisfy Eq. (3.10) and therefore correspond to representations of the two-colour Temperley–Lieb algebra. For more general representations of this algebra we refer to Ref. [10].

### 4.1 Representation space

To give our representations of the two-colour BWM algebra we first have to define our space of states.

#### 4.1.1 Local states

First we define a *local state* $a$. This is an $(n_1+n_2)$-dimensional vector

$$a = \sum_{c=1,2} \sum_{i=1}^{n_c} a_i^{(c)} \epsilon_i^{(c)}, \tag{4.1}$$

with $\{\epsilon_i^{(c)}\}_{i=1,\ldots,n_c}^{c=1,2}$ a set of orthonormal vectors,

$$(\epsilon_i^{(c)}, \epsilon_j^{(c')}) = \delta_{i,j} \, \delta_{c,c'} . \tag{4.2}$$

The entries $a_i^{(c)}$ of $a$ satisfy restrictions defined as follows.

1. Choose a pair $(\mathcal{A}_1, \mathcal{A}_2)$ with $\mathcal{A}_c = $ B,C,D.

2. Set

$$L_c = t_c(\ell_c + g_c), \tag{4.3}$$

with $\ell_c \in \mathbb{Z}_{>0}$ fixed but arbitrary and with $t_c$ and $g_c$ given in Table 1. The integers $n_1$ and $n_2$, labelling the dimension of a local state vector $a$, are also to be fixed, and, depending on the choice of $\mathcal{A}_c$, must satisfy: $n_c \geq 2$ if $\mathcal{A}_c = $ B, $n_c \geq 1$ if $\mathcal{A}_c = $ C and $n_c \geq 3$ if $\mathcal{A}_c = $ D.

3. If $\mathcal{A}_c = $ B:

$$0 < a_{n_c}^{(c)} < \ldots < a_2^{(c)} < a_1^{(c)}, \quad a_1^{(c)} + a_2^{(c)} < L_c$$

$$a_1^{(c)}, \ldots, a_{n_c}^{(c)} \in \mathbb{Z} \quad \text{or} \quad a_1^{(c)}, \ldots, a_{n_c}^{(c)} \in \mathbb{Z} + \frac{1}{2}. \tag{4.4}$$

If $\mathcal{A}_c = $ C:

$$0 < a_{n_c}^{(c)} < \ldots < a_2^{(c)} < a_1^{(c)} < L_c/2$$

$$a_1^{(c)}, \ldots, a_{n_c}^{(c)} \in \mathbb{Z}. \tag{4.5}$$

If $\mathcal{A}_c = $ D:

$$0 < a_{n_c}^{(c)} < \ldots < a_2^{(c)} < a_1^{(c)}, \quad 0 < a_{n_c}^{(c)} + a_{n_c-1}^{(c)}, \quad a_1^{(c)} + a_2^{(c)} < L_c$$

$$a_1^{(c)}, \ldots, a_{n_c}^{(c)} \in \mathbb{Z} \quad \text{or} \quad a_1^{(c)}, \ldots, a_{n_c}^{(c)} \in \mathbb{Z} + \frac{1}{2}. \tag{4.6}$$

We remark that the symbols B, C, and D used to label the different possible choices for $\mathcal{A}_c$ reflect the underlying Lie algebraic structure. In particular, the sets $\{a_i^{(c)}\}_{i=1}^{n_c}$ defined by Eqs. (4.4)–(4.6) are in one-to-one correspondence with the level $\ell_c$ dominant integral weights of the respective affine algebra $(\mathcal{A}_c)_n^{(1)}$.

| $\mathcal{A}_c$ | $g_c$ | $t_c$ | $\eta_c$ | $\sigma_c$ | $h_c(a)$ |
|---|---|---|---|---|---|
| B | $2n_c - 1$ | 1 | $n_c - \frac{1}{2}$ | 1 | $\sin(a\lambda_c)$ |
| C | $n_c + 1$ | 2 | $n_c + 1$ | $-1$ | $\sin(2a\lambda_c)$ |
| D | $2n_c - 2$ | 1 | $n_c - 1$ | 1 | 1 |

Table 1

### 4.1.2 Admissibility rules

In the following it will be convenient to extend the subscript of $\epsilon_i^{(c)}$ by setting $\epsilon_{-i}^{(c)} = -\epsilon_i^{(c)}$. Using this notation we define two local states $a$ and $b$ to be *adjacent* if $a - b = \epsilon_\mu^{(c)}$, for some $\mu \in \{\pm 1, \ldots, \pm n_c\}$, and some $c \in \{1, 2\}$. If we draw the set of all local states as a collection of nodes, we can represent the adjacency of two nodes $a$ and $b$ graphically, by drawing a bond between $a$ and $b$, see Figure 2a.

We can now introduce the notion of a *path* as an ordered sequence of adjacent local states, i.e., a path is a sequence of the type

$$a, a + \epsilon_\mu^{(c')}, a + \epsilon_\mu^{(c')} + \epsilon_\nu^{(c'')}, \ldots, \tag{4.7}$$

with $\mu \in \{\pm 1, \ldots, \pm n_{c'}\}$, $c' \in \{1, 2\}$ and $\nu \in \{\pm 1, \ldots, \pm n_{c''}\}$, $c'' \in \{1, 2\}$, etc. Graphically one can think of a path as a sequence of steps along bonds connecting adjacent local states.

If $\mathcal{A}_c = B$, we further extend the subscript of $\epsilon_i^{(c)}$ by introducing the symbol $\epsilon_0^{(c)}$. In the $(n_1 + n_2)$-dimensional space of local state vectors, $\epsilon_0^{(c)}$ corresponds to the zero vector. Hence, $a + \epsilon_i^{(c)} = a$. If we extend our notion of a path by also allowing $\mu = 0$ if $c' = c$ and $\nu = 0$ if $c'' = c$, etc, in Eq. (4.7), we can represent the step from node $a$ to node $a + \epsilon_0^{(c)} = a$ by a step along a tadpole, see figure 2b.

We wish for this rather clumsy graphical way of denoting the zero vector because we can have the situation $(\mathcal{A}_1, \mathcal{A}_2) = (B, B)$. In this case we have both the symbols $\epsilon_0^{(1)}$ and $\epsilon_0^{(2)}$. Although they both correspond to the zero vector in the local state space, we do have to distinguish $a + \epsilon_0^{(1)}$ and $a + \epsilon_0^{(2)}$. With the previous graphical notation this can be made clear by drawing two tadpoles on $a$, see figure 2c. A step "along" $\epsilon_0^{(1)}$ corresponds to a step from $a$ to $a$ along the tadpole labelled 1, and a step along $\epsilon_0^{(2)}$ corresponds to a step from $a$ to $a$ along the tadpole labelled 2.

With the above we now define an *admissible path* $\{a\}$ as a sequence of $N + 2$ local states

$$\{a\} = a, a + \epsilon_{\mu_1}^{(c_1)}, a + \epsilon_{\mu_1}^{(c_1)} + \epsilon_{\mu_2}^{(c_2)}, a + \epsilon_{\mu_1}^{(c_1)} + \epsilon_{\mu_2}^{(c_2)} + \epsilon_{\mu_3}^{(c_3)}, \ldots \equiv a_0, a_1, \ldots, a_{N+1}, \tag{4.8}$$

subject to the restrictions:

1. $c_j \in \{1, 2\}$.

2. $\mu_j = \{\pm 1, \ldots, \pm n_{c_j}\}$ if $\mathcal{A}_{c_j} = C, D$.

3. $\mu_j = \{0, \pm 1, \ldots, \pm n_{c_j}\}$ if $\mathcal{A}_{c_j} = B$,

4. $\mu_j \neq 0$ if $(a + \epsilon_{\mu_1}^{(c_1)} + \ldots + \epsilon_{\mu_{j-1}}^{(c_{j-1})}, \epsilon_{n_{c_j}}^{(c_j)}) = \frac{1}{2}$.

The space $\mathcal{P}_N$ of all admissible paths $\{a\}$ will be the representation space of our two-colour BWM algebra.

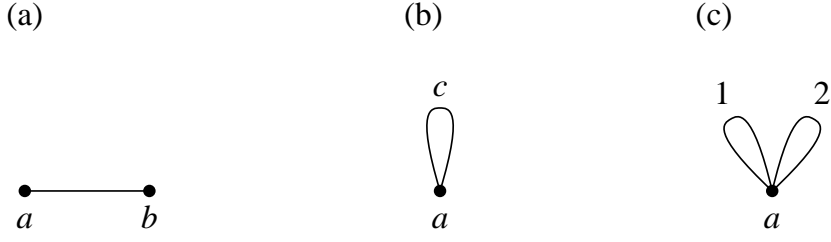

Figure 2: (a) Graphical representations of two adjacent local states $a$ and $b$. (b) A tadpole representing a step from $a$ to $a + \epsilon_0^{(c)} = a$. (c) Tadpoles representing steps from $a$ to $a + \epsilon_0^{(1)} = a$ and $a + \epsilon_0^{(2)} = a$, respectively.

## 4.2 Representations labelled by $(\mathcal{A}_1, \mathcal{A}_2)$

Now that we have defined $\mathcal{P}_N$, we can give the actual matrix elements of our representations labelled by the pair $(\mathcal{A}_1, \mathcal{A}_2)$. Since all the two-colour BWM generators act non-trivially at positions $j$ and $j+1$ only, we set

$$(\mathcal{O}_j)_{\{a\},\{b\}} = \mathcal{O}\begin{pmatrix} & b_j & \\ a_{j-1} & & a_{j+1} \\ & a_j & \end{pmatrix} \prod_{k \neq j} \delta_{a_k, b_k}, \qquad (4.9)$$

with $\mathcal{O}_j$ any of the operators in (2.1).

We define two more variables

$$\lambda_c = \frac{s_c \pi}{L_c}, \qquad c = 1, 2, \qquad (4.10)$$

with $L_c$ defined in (4.3) and with $s_c \in \mathbb{Z}$ arbitrary but coprime with $L_c$. We then have representations of the two-colour BWM algebra with constants fixed by

$$\begin{aligned} q_c &= e^{-i\lambda_c} \\ \omega_c &= \sigma_c e^{-i(2\eta_c + \sigma_c)\lambda_c} \\ \sqrt{Q_c} &= \frac{\sin[(1 + \eta_c \sigma_c)\lambda_c] \sin 2\eta_c \lambda_c}{\sin \lambda_c \sin \eta_c \lambda_c} \end{aligned} \qquad (4.11)$$

and with non-zero matrix elements given by

$$p^{(c,c')}\begin{pmatrix} & a+\epsilon_\mu^{(c)} & \\ a & & a+\epsilon_\mu^{(c)}+\epsilon_\nu^{(c')} \\ & a+\epsilon_\mu^{(c)} & \end{pmatrix} = 1$$

$$e^{(c,c')}\begin{pmatrix} & a+\epsilon_\nu^{(c')} & \\ a & & a \\ & a+\epsilon_\mu^{(c)} & \end{pmatrix} = \left(G_{a,\mu}^{(c)} G_{a,\nu}^{(c')}\right)^{1/2}$$

$$b^{\pm(c,c)}\begin{pmatrix} & a+\epsilon_\mu^{(c)} & \\ a & & a+2\epsilon_\mu^{(c)} \\ & a+\epsilon_\mu^{(c)} & \end{pmatrix} = e^{\pm i\lambda_c} \qquad \mu \neq 0$$

$$b^{\pm(c,c)}\begin{pmatrix} & a+\epsilon_\mu^{(c)} & \\ a & & a+\epsilon_\mu^{(c)}+\epsilon_\nu^{(c)} \\ & a+\epsilon_\mu^{(c)} & \end{pmatrix} = -\mathrm{e}^{\mp\mathrm{i}(a_\mu^{(c)}-a_\nu^{(c)})\lambda_c}\frac{\sin\lambda_c}{\sin[(a_\mu^{(c)}-a_\nu^{(c)})\lambda_c]} \qquad \mu\neq\pm\nu$$

$$b^{\pm(c,c)}\begin{pmatrix} & a+\epsilon_\nu^{(c)} & \\ a & & a+\epsilon_\mu^{(c)}+\epsilon_\nu^{(c)} \\ & a+\epsilon_\mu^{(c)} & \end{pmatrix}$$
$$= -\left(\frac{\sin[(a_\mu^{(c)}-a_\nu^{(c)}+1)\lambda_c]\,\sin[(a_\mu^{(c)}-a_\nu^{(c)}-1)\lambda_c]}{\sin^2[(a_\mu^{(c)}-a_\nu^{(c)})\lambda_c]}\right)^{1/2} \qquad \mu\neq\pm\nu$$

$$b^{\pm(c,c)}\begin{pmatrix} & a+\epsilon_\nu^{(c)} & \\ a & & a \\ & a+\epsilon_\mu^{(c)} & \end{pmatrix} = \left(G_{a,\mu}^{(c)}G_{a,\nu}^{(c)}\right)^{1/2}\mathrm{e}^{\mp\mathrm{i}(a_\mu^{(c)}+a_\nu^{(c)}+1)\lambda_c}\frac{\sin\lambda_c}{\sin[(a_\mu^{(c)}+a_\nu^{(c)}+1)\lambda_c]} \qquad \mu\neq\nu$$

$$b^{\pm(c,c)}\begin{pmatrix} & a+\epsilon_\mu^{(c)} & \\ a & & a \\ & a+\epsilon_\mu^{(c)} & \end{pmatrix} = \left(G_{a,\mu}^{(c)}-1\right)\mathrm{e}^{\mp\mathrm{i}(2a_\mu^{(c)}+1)\lambda_c}\frac{\sin\lambda_c}{\sin[(2a_\mu^{(c)}+1)\lambda_c]} \qquad \mu\neq 0$$
$$= \left(1-H_{a,\mu}^{(c)}\right)\mathrm{e}^{\mp\mathrm{i}(2a_\mu^{(c)}+1)\lambda_c}\frac{\sin\lambda_c}{\sin[(2a_\mu^{(c)}-2\eta_c+1)\lambda_c]}$$

$$b^{(c,\bar c)}\begin{pmatrix} & a+\epsilon_\nu^{(\bar c)} & \\ a & & a+\epsilon_\mu^{(c)}+\epsilon_\nu^{(\bar c)} \\ & a+\epsilon_\mu^{(c)} & \end{pmatrix} = 1\,, \tag{4.12}$$

with $a_0^{(c)}=-1/2$ and with functions $G_{a,\mu}^{(c)}$ and $H_{a,\mu}^{(c)}$ given by

$$G_{a,\mu}^{(c)} = \sigma_c\,\frac{h_c(a_\mu^{(c)}+1)}{h_c(a_\mu^{(c)})}\prod_{\substack{|\nu|=1 \\ \nu\neq\pm\mu}}^{n_c}\frac{\sin[(a_\mu^{(c)}-a_\nu^{(c)}+1)\lambda_c]}{\sin[(a_\mu^{(c)}-a_\nu^{(c)})\lambda_c]} \qquad \mu\neq 0$$

$$G_{a,0}^{(c)} = 1$$

$$H_{a,\mu}^{(c)} = \sum_{\substack{|\nu|=1 \\ \nu\neq\mu}}^{n_c} G_{a,\nu}^{(c)}\,\frac{\sin[(a_\mu^{(c)}+a_\nu^{(c)}-2\eta_c+1)\lambda_c]}{\sin[a(_\mu^{(c)}+a_\nu^{(c)}+1)\lambda_c]}\,. \tag{4.13}$$

The functions $h_c$ and the constants $\sigma_c$ and $\eta_c$ in the above equations are listed in Table 1.

The proof that Eqs. (4.9)–(4.13) indeed provide representations of the two-colour BWM algebra is straightforward. All relations in Eqs. (2.4)–(2.8) which involve a single colour are satisfied because our representations are constructed from representations of the original (one-colour) BWM algebra as given by Deguchi et al. [15]. Any relation involving both colours holds trivially using the factorization property of the Temperley–Lieb operators $e^{(c,c')}$ and the simple form of the mixed braids $b^{(c,\bar c)}$ in Eq. (4.12).

## 4.3 Solvable RSOS models

We now return to the Baxterization (3.5). From Eq. (3.3), we see that in order to obtain a solution of the YBE we require

$$\lambda_1 = \lambda_2 = \lambda \tag{4.14}$$

for our representations of the two-colour algebra given in Sec. 4.2. This in turn implies $s_1/L_1 = s_2/L_2$, see (4.10). Nevertheless, we still can use representations with different values of $\eta_c$ and $\sigma_c$, which determine the value of $\eta$ in Eq. (3.5) by

$$\eta = \eta_1 + \eta_2 + \frac{\sigma_1 + \sigma_2}{2} \pm \frac{\pi(\sigma_1 - \sigma_2)}{2\lambda} \tag{4.15}$$

as is easily seen from Eqs. (3.4) and (4.11). For any such representation, Eq. (3.5) gives rise to a solvable lattice model of RSOS type. The Boltzmann weights

$$W \left( a \begin{array}{c} d \\ c \\ b \end{array} \middle| u \right) \; = \; a \diamondsuit_{b}^{d} c \quad (u) \tag{4.16}$$

are given by the matrix elements of the face operators [2]

$$\left( X_j(u) \right)_{\{a\},\{b\}} = W \left( a_{j-1} \begin{array}{c} b_j \\ a_{j+1} \\ a_j \end{array} \middle| u \right) \prod_{k \neq j} \delta_{a_k, b_k} . \tag{4.17}$$

To our knowledge, apart from the models labelled by $(C,C)$ with $n_1 = n_2 = 1$ [10, 16] these models are new. The corresponding adjacency graphs are given by products (in the sense of Sec. 4) of the graphs underlying the $B_n^{(1)}$, $C_n^{(1)}$ and $D_n^{(1)}$ models of Ref. [17].

# 5  Discussion

In this paper, we have constructed solvable RSOS models based on a two-colour generalization of the BWM algebra. In particular, we have presented a Baxterization of the two-colour algebra. This ensures that any suitable representation of the algebra gives rise to a solvable lattice model. Representations of the two-colour BWM algebra are constructed from any pair $(\mathcal{A}_1, \mathcal{A}_2)$, where $\mathcal{A}_c$ denotes an RSOS-type representation of the ordinary BWM algebra labelled by either the $B_n^{(1)}$, $C_n^{(1)}$, or $D_n^{(1)}$ affine Lie algebra. We have also shown that the known Baxterizations of the dilute BWM algebra and of the two-colour Temperley–Lieb algebra are contained in our Baxterization of the two-colour BWM algebra as special cases.

Although we have restricted ourselves to the RSOS type representations, it is straightforward to include vertex-type representations. If both $\mathcal{A}_1$ and $\mathcal{A}_2$ are of vertex-type, one obtains solvable vertex models. Combining a vertex-type and an RSOS-type representation results in mixed RSOS-vertex models.

Finally, we mention that it would of course be interesting to generalize our work to an arbitrary number of colours. Clearly, our method of constructing representations of the two-colour algebra can be applied to yield representations of a multi-colour BWM algebra. However, we have not succeeded in finding a Baxterization beyond the case of two colours.

# Acknowledgements

We thank Paul Pearce for helpful discussions. This work has been supported by the Samenwerkingsverband FOM/SMC Mathematische Fysica and the Australian Research Council.